\documentclass{ws-rv9x6}
\usepackage{ws-rv-van}             
\begin{document}
\bibliographystyle{ws-rv-van}

\chapter{Feshbach resonances in ultracold gases}

\author[S.J.J.M.F. Kokkelmans]{Servaas Kokkelmans}
\address{Department of physics, Eindhoven University of Technology, P.O. Box 513, 5600 MB Eindhoven, The Netherlands}



\begin{abstract}
In this chapter, we describe scattering resonance phenomena in general, and focus on the mechanism of Feshbach resonances, for which a multi-channel treatment is required. We derive the dependence of the scattering phase shift on magnetic field and collision energy.  From this, the scattering length and effective range coefficient can be extracted, expressions which are particularly useful for ultracold gases.
\end{abstract}

\body

\section{Introduction}

One of the most dramatic phenomena in quantum scattering physics \cite{Taylor72} is the occurance of resonances. They often lead to peaks in elastic cross sections as function of energy or magnetic field, but also to large increments in loss rates. There are several different types of resonance phenomena, but they all have in common that their origin lies in a nearby bound state, which is often a long-lived state with a corresponding width, and will therefore eventually decay. The resonance may also be a result of a state which is almost bound, and to separate such states from real bound states they are given the distinctive name 'virtual state'. This close connection between bound states and scattering resonances becomes apparent as bound states correspond to poles in the scattering matrix. 

A {\it shape resonance} \cite{Sakurai93,Kukulin89} is probably the most clear example of a scattering resonance, where a long-lived state arises within a potential barrier. Most commonly this is a centrifugal barrier resulting from scattering with non-zero angular momentum ($l>0$). More details on the atom-atom interaction, internal state and angular momentum quantum numbers will be given in the next section. The collision energy acts as a control parameter which allows the system to tune the cross section. A general feature of these resonances is that the resonant contribution to the scattering phase shift $\delta_l$ changes rapidly from zero to $\pi$ on an energy interval characterized by a width $\Gamma$ around the resonance position $E_R$:
\begin{equation}
 \tan \delta_l(k)=\frac{\Gamma/2}{E-E_R}.
\end{equation}
Here $E=\hbar^2 k^2/2m$ with $m$ the reduced mass\footnote{More commonly the reduced mass is indicated by $\mu$, however, to avoid confusion with magnetic moments which are also indicated by $\mu$, we use $m$ instead.}. When background scattering processes are ignored, this leads to the cross section
\begin{equation}
 \sigma_l(k)=\frac{4 \pi (2l+1)}{k^2} \sin^2 \delta_l (k) = \frac{4 \pi (2l+1)}{k^2} \frac{\Gamma^2/4}{(E-E_R)^2+\Gamma^2/4},
\label{crossection}
\end{equation}
which is the famous Breit-Wigner formula. Note that for identical particles in the same internal state, these cross sections are a factor 2 larger.

A {\it potential resonance} \cite{Taylor72,Kukulin89} occurs in the absence of a barrier, and is therefore a purely $s$-wave phenomenon ($l=0$). This resonance is a result of a bound or virtual state close to the collision threshold of a single-channel interaction potential, and is characterised by a scattering length $a_S$ which is much larger than the range of the interaction potential. Because of the single-channel nature of the interaction, this resonance is generally not easily tunable. In ultracold gases, the $s$-wave scattering length is an important quantity to characterize the binary interactions, and is related to the $s$-wave cross section of Eq.~(\ref{crossection}) via $\sigma_0=4 \pi a_S^2$.

For a {\it Feshbach resonance} \cite{Kukulin89,Feshbach92}, it is essential that more than one collision channel is present, and therefore it is quite different from the above-mentioned resonances which are single-channel phenomena. In ultracold collisions these channels correspond to different spin configurations of the atomic pair, which give rise to tunable internal state energies via external (magnetic) fields. In essence the resonance is an interference of a background scattering processes in the incoming (energetically open) channel, with a resonant scattering process in an energetically closed channel. Therefore a transition must be made of the open channel to the closed channel, and the result is that a closed-channel bound state transforms into a long-lived resonant state. This will be explained in detail in the next sections.

Feshbach resonances are in ultracold atomic physics mostly appreciated for the tuning possibilities of the scattering length via external magnetic fields. Here the Zeeman effect allows for relative changes in the internal state energies, and makes it possible to tune a closed-channel bound state in or out resonance. The characteristic dispersive shape of the scattering length is given by
\begin{equation}
 a_S=a_{\rm bg} \left(1-\frac{\Delta B}{B-B_0} \right).
\label{aofB}
\end{equation}
Here is $a_{\rm bg}$ the background scattering length, a result of the background collision in the open channel, $B_0$ is the field of resonance  and $\Delta B$ the field range over which the scattering length changes sign. The treatment of this resonance phenomenon by Feshbach took place in nuclear physics \cite{Feshbach58,Feshbach62}, and the term Feshbach resonance only appeared in the context of quantum gases in 1976 as an enhanced loss mechanism, which should be avoided \cite{stwalley76,chin10}. Later in 1993 it was found that Feshbach resonances can be used to change the sign and strength of the interatomic interactions \cite{tiesinga93,chin10}, something which was not realized before, and only five years later these tuning possibilities of Feshbach resonances where first observed in ultracold gases \cite{inouye98,courteille98}. The use of magnetic Feshbach resonances is now common practise in ultracold gases experiments. The possibility of a tunable scattering length in fermionic gases made it possible to explore the physics of the BCS-BEC crossover (see chapter 9). Tunable interactions also play a role in many other experiments, for instance in the study of Efimov three-body physics, where some observables of the system show log-periodic behavior in $a_S$. Feshbach resonances also facilitate the creation of ground-state molecules starting from ultacold atoms \cite{Kokkelmans01,Lang08,Ni08}.

A {\it Fano resonance} \cite{Fano61} is essentially equivalent to a Feshbach resonance, but the term Fano resonance is usually associated to the asymmetric lineshape of the cross section (in contrast to the symmetric Breit-Wigner resonance) as function of the energy. While the Feshbach resonance is usually associated to the magnetic field tuning of the scattering length, their origin is the same: it is a result of an interference between a background and a resonant scattering process.

In the following sections, we focus on the description of Feshbach resonances, and first start with a treatment of the underlying interatomic interactions, which are specifically tailored towards a system of homonuclear alkali-metal atoms. Then we introduce the concept of multi-channel scattering, and derive the coupled-channels radial Schr\"odinger equation. After that, we describe the projection operator formalism of Feshbach resonances which lead to analytic expressions for the scattering matrix which include physical parameters such as the background scattering length, the width and position of the resonance. Here a distinction is made between Feshbach resonances with and without resonant open channel scattering. The full treatment that allows for resonant open channel interactions requires the introduction of an additional bound or virtual state in the open channel. Finally some useful expressions are derived for the different types of resonances, relevant for ultracold scattering, such as the energy-dependent scattering phase shift, the scattering length and the effective range coefficient.

\section{Atom-atom interactions}

For a given external magnetic field, the effective two-body Hamiltonian takes the following form:
\begin{equation} \label{Hamiltonian}
H = H_0+V,
\end{equation}
with the relative kinetic energy operator and single-particle interactions put together in
\begin{equation}
 H_0=\frac{{\bf p}^{2}}{2m} + \sum^{2}_{j=1} \left( V^{\textrm{hf}}_{j} + V^{\textrm{Z}}_{j}\right),
 \label{single_particle_ham}
\end{equation}
with $m$ the reduced mass, and
\begin{equation}
 V= V^{\rm cen} + V^{\rm dd}_{\rm spin}
\label{two_particle_ham}
\end{equation}
the two-body interaction. The single-particle interactions in Eq.~(\ref{single_particle_ham}) describe the hyperfine and Zeeman energies of the individual atoms $j=1,2$, which are given explicitly by
\begin{equation}
V^{\textrm{hf}}_{j} = \frac{a^{\textrm{hf}}_{j}}{\hbar^2} {\bf s}_{j} \cdot {\bf i}_{j},
\end{equation}
where ${\bf s}_{j}$ and ${\bf i}_{j}$ are the electron and nuclear spin operators of atom $j$ and the constant $a^{\textrm{hf}}_{j}$ is related to the hyperfine splitting, and
\begin{equation}
    V^{\textrm{Z}}_{j} = (\gamma_{e} {\bf s}_{j} - \gamma_{N} {\bf i}_{j}) \cdot {\bf B},
    \label{ZeemanInt}
\end{equation}
where $\gamma_{e}$ and $\gamma_{N}$ are the electronic and nuclear gyromagnetic ratios, respectively. Note that the valence electron feels a small effect from the electrons in the inner shells, and $\gamma_{e}$ differs slightly from the gyromagnetic ratio of a free electron. The ratio between $\gamma_{e}$ and $\gamma_{N}$ is on the order of $10^3$. The magnetic field $\bf B$ is assumed to be constant over the range of the interaction, which is typically on the order of $50-100$ ${\rm a_0}$ for the alkali atoms, where ${\rm a_0} \approx 5.2917 \times 10^{-11}$ m is the Bohr radius.

\begin{figure}
\begin{center}
\includegraphics[width=6.5cm]{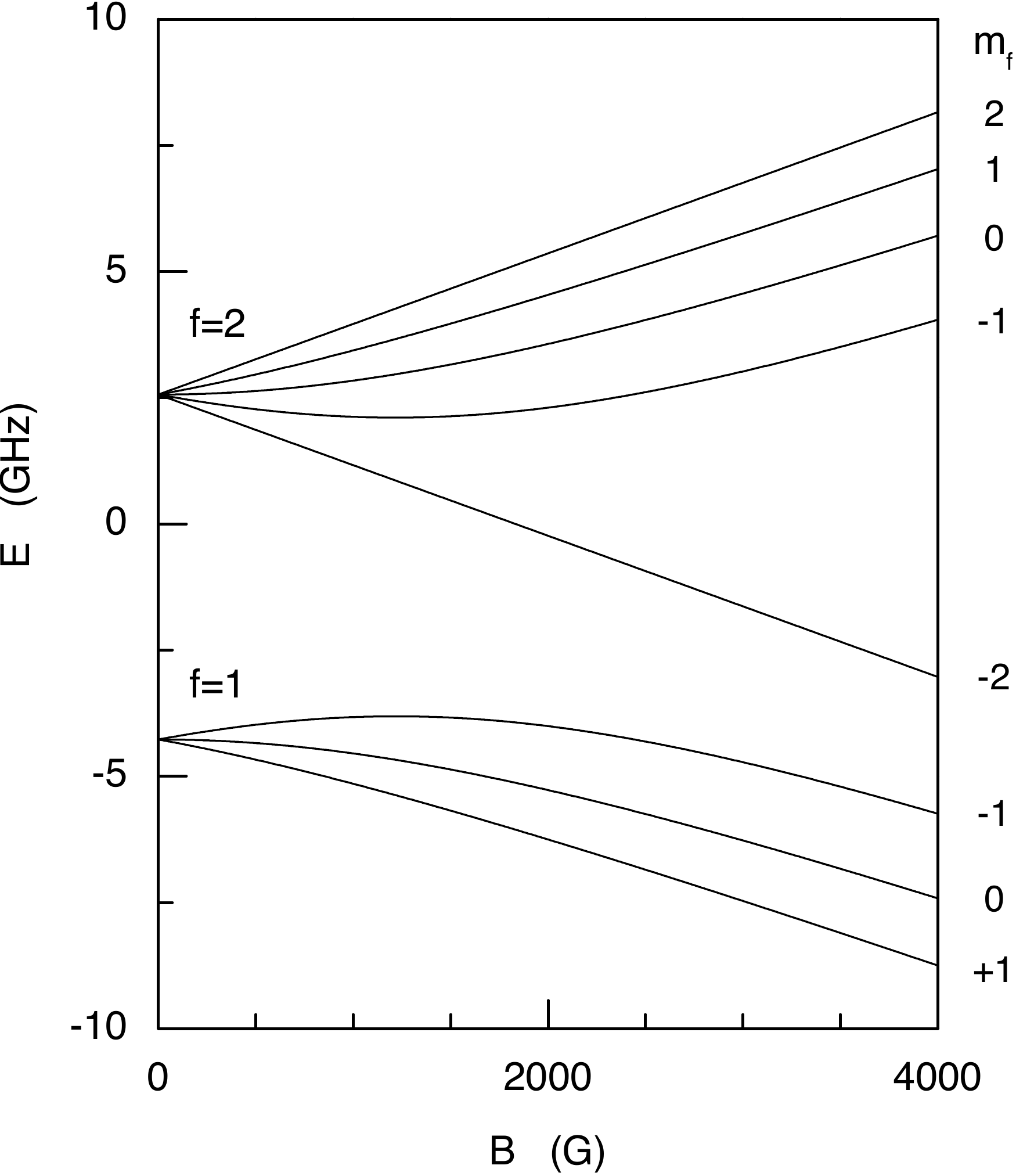}
\caption{Single atom hyperfine diagram for $^{87}$Rb. These eigenvalues $\varepsilon_\alpha$ of the combined hyperfine and Zeeman interaction are often denoted by quantum numbers $f$ and $m_f$, even though $f$ is a good quantum number for $B=0$ only.} \label{hyperfine}
\end{center}
\end{figure}

At large separations, the interatomic interactions are negligible and the two-atom system can be described by tensor products of the single-atom eigenstates, which we call the asymptotic states. At zero magnetic field, the electronic and nuclear spin combine to an effective spin ${\bf f=s+i}$ and the single-atom hyperfine states can be labeled by $|fm_f\rangle \equiv |\alpha \rangle$. Although $f$ is strictly speaking not a good quantum number at non-zero magnetic fields, it will be used to label the asymptotic (channel) states at all magnetic fields. The hyperfine and Zeeman interactions lead to the well known dependence of the energy $\varepsilon_\alpha$ of the hyperfine states $|\alpha \rangle$ on the magnetic field, an example of which is given in Fig.~\ref{hyperfine} for $^{87}$Rb.

The first term of the two-body interaction in Eq.~(\ref{two_particle_ham}) describes the central interatomic interaction $V^{\rm cen}$, and represents all Coulomb interactions between the electrons and nuclei of both atoms. The central interaction does only depend on the distance between the nuclei $r\equiv |{\bf r}|$ and is invariant under rotations of the orbital system. As a consequence, the orbital angular momentum vector $\bf l$ is conserved and, choosing the direction of the magnetic field as the quantization axis, $l$ and $m_l$ are good quantum numbers. Simultaneously, the projection $m_F$ of the total spin $\bf F=S+I$ along this axis is conserved during the collision. As usual, ${\bf S=s_{1}+s_{2}}$ is the total electron spin and ${\bf I=i_{1}+i_{2}}$ is the total nuclear spin. In the presence of only the internal and central interactions this leads to the selection rules $\Delta m_F=0$, $\Delta l=0$ and $\Delta m_l$=0.

\begin{figure}
\begin{center}
\includegraphics[width=8cm]{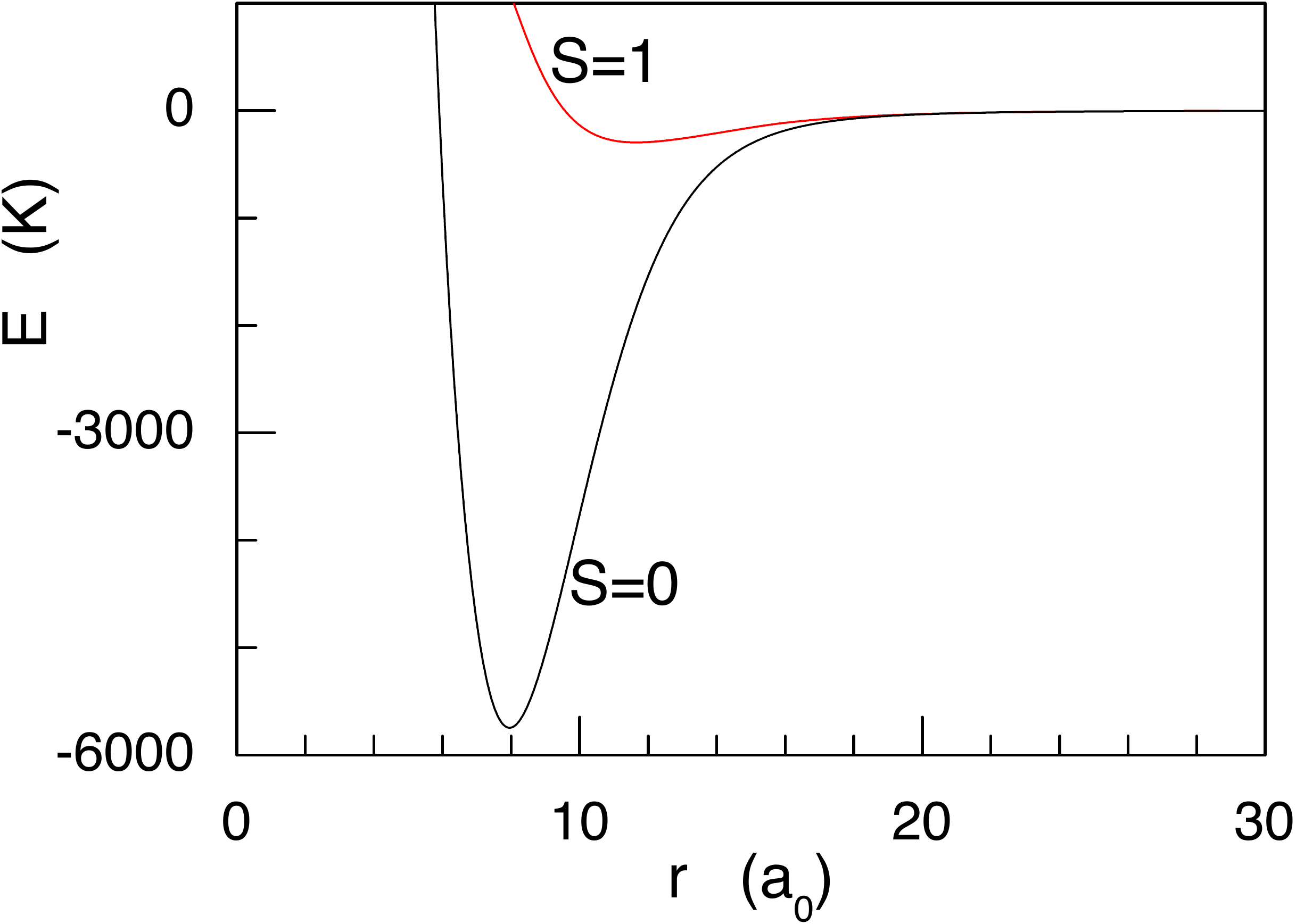}
\caption{Singlet ($S=0$) and triplet ($S=1$) potentials for ground-state rubidium.} \label{groundRb}
\end{center}
\end{figure}

The central interaction conserves the total spin ${\bf F=S+I}$, as well as $\bf S$ and $\bf I$ separately and $V^{\rm cen}$ can be decomposed into a singlet and triplet term
\begin{equation}
    V^{\rm cen}(r) = V_{S}(r)P_{S} + V_{T}(r)P_{T},
\end{equation}
where $P_{S}$ and $P_{T}$ are the projection operators on the singlet ($S=0$) and triplet ($S=1$) subspaces. The asymptotic singlet and triplet potentials are given by
\begin{equation}
V_{S,T}(r) = V_{\rm disp}(r) - (-1)^{S} V_{\rm ex}(r),
\end{equation}
with the familiar dispersive form at long range
\begin{equation}
V^{\rm disp}(r) =-\frac{C_6}{r^6}-\frac{C_8}{r^8}-\frac{C_{10}}{r^{10}} - \ldots,
\end{equation}
where $C_n$ are the van der Waals dispersion coefficients. The exchange potential is related to the symmetry of the electronic wave function and can be given by an analytic (asymptotic) expression \cite{Smirnov65}
\begin{equation}
    V_{\textrm{ex}}(r) = Jr^{\frac{7}{2\alpha}-1}e^{-2\alpha r},
\end{equation}
where $\alpha$ is related to the atomic ionization energy as $-\alpha^{2}/2$ and $J$ is a normalization constant. Note that $\alpha$ and $r$ are given in atomic units and are dimensionless. The ground-state potentials for rubidium are shown in Fig.~\ref{groundRb}.

The second term in Eq.~(\ref{two_particle_ham}) describes the non-central anisotropic part of the interactions. It contains the direct dipole-dipole interaction of the electronic spins \cite{Moerdijk95} and the second-order spin-orbit coupling \cite{chin10}. The magnetic dipole moment of atom $j$ is
\begin{equation}
{\boldsymbol \mu_j} = \mu {\boldsymbol \sigma_j},
\end{equation}
with $\mu$ the electronic magnetic dipole moment and $\boldsymbol \sigma_j$ the Pauli-spin vector of the valence electron of atom $j$. The direct spin-spin interaction then takes the form:
\begin{equation}\label{spinspin}
V^{\rm dd}_{\rm spin} = \frac{\mu^2}{4\pi \mu_0^{-1}r^3} \left[ {\boldsymbol \sigma_1 \cdot \boldsymbol \sigma_2} - 3\left( {\boldsymbol \sigma_1 \cdot \bf \hat{r} } \right)\left( {\boldsymbol \sigma_2 \cdot \bf \hat{r} } \right) \right],
\end{equation}
where $\mu_0$ is the magnetic constant. The second-order spin-orbit term has the same angular dependence, but a different $r$-dependent prefactor. The spin-spin interaction is rather weak and in many cases it can be neglected. However, in $^{87}$Rb and $^{133}$Cs the spin-spin interaction gives rise to Feshbach resonances by coupling $l=2$ and even $l=4$ (quasi-)bound states to the s-wave entrance channel, and these resonances have been observed experimentally \cite{marte02,chin04}.

The angle-dependent structure of the spin-spin anisotropic interactions can be rewritten in terms of a scalar product of two irreducible spherical tensors of rank 2. Due to this nature, the anisotropic part of the interaction obeys triangle type selection rules and couples states with different orbital angular momentum according to the selection rule $l-l'=0,2$, with the exception of $l=0 \rightarrow l'=0$ which is forbidden \cite{messiah65}.

In the presence of these anisotropic interactions, the two-atom system is not invariant under independent rotations of the orbital and spin systems and the projections $m_F$ and $m_l$ are not separately conserved. However, the anisotropic spin-spin interactions are invariant under simultaneous 3D rotations of the internuclear vector $\bf r$ and the spins. The total angular momentum is thus conserved. In the presence of a magnetic field, the Zeeman term breaks the rotational symmetry and introduces a cylindrical symmetry. Therefore, only the projection of the total angular momentum along the magnetic field axis will be conserved. Consequently, in the presence of a magnetic field the selection rule becomes $\Delta m_{\rm tot} =\Delta (m_F+m_l)=0$.

\section{Coupled-channels equation}

\begin{figure}
\begin{center}
\includegraphics[width=8cm]{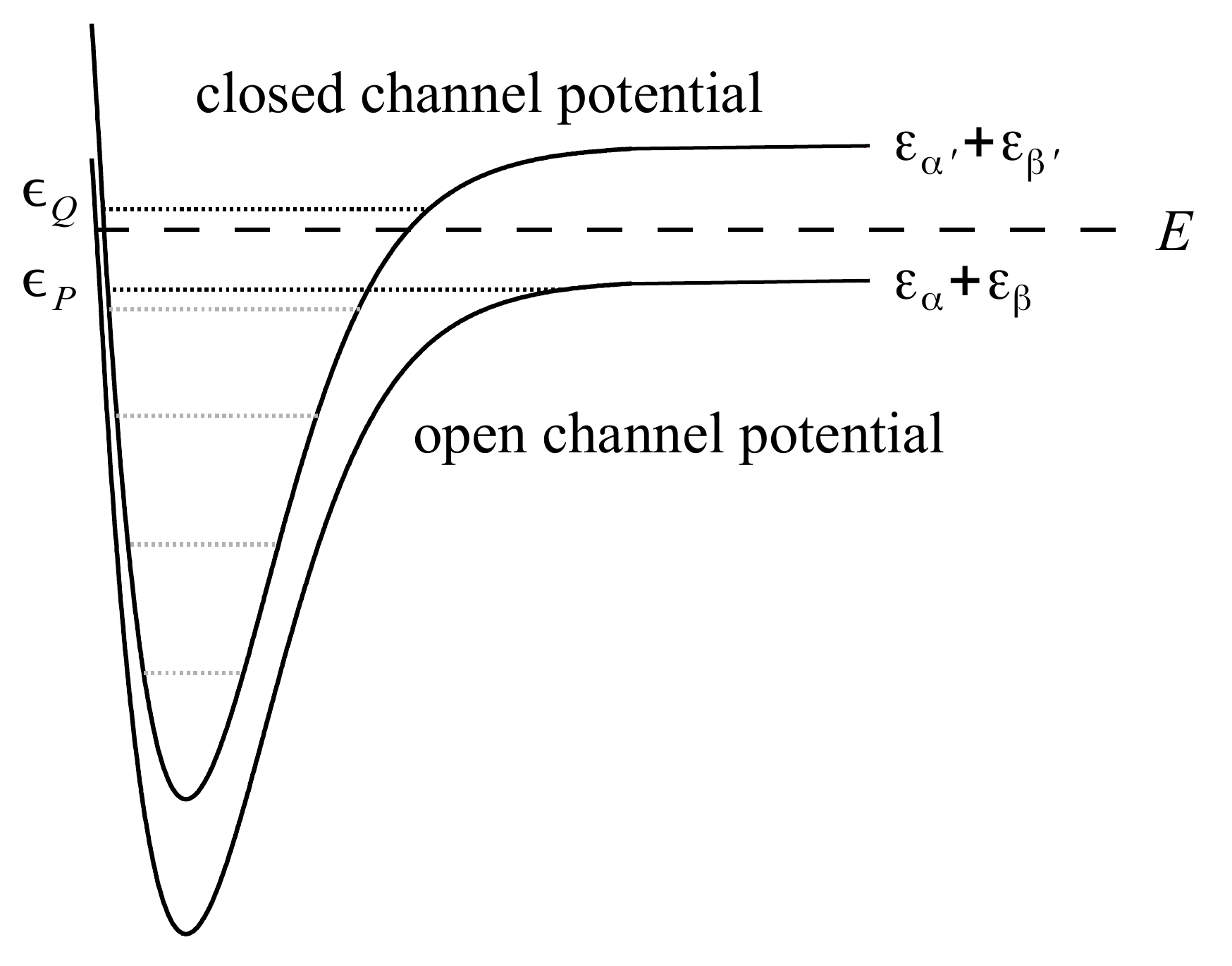}
\caption{Schematic representation of two coupled channels potentials. The asymptotic energy of the open channel potential, $\varepsilon_\alpha+\varepsilon_\beta$, and closed channel potential, $\varepsilon_\alpha'+\varepsilon_\beta'$, are the sum of the single-particle internal energies in that particular channel. Two atoms always start a collision in an open channel, which means that the energy $E$ is larger than the asymptotic energy of that channel. This asymptotic energy is referred to as the collision threshold, which in this example is given by $\varepsilon_{\rm thr}=\varepsilon_\alpha+\varepsilon_\beta$. At short distance transitions are allowed to an energetically closed channel, from which the atoms cannot escape. Also indicated are the highest bound state in the open channel with binding energy $\epsilon_P$, and several bound states in the closed channel potential where a particular one, with binding energy $\epsilon_Q$, is responsible for a Feshbach resonance.} \label{Feshbach_mechamism}
\end{center}
\end{figure}

One can think of a collision process dynamically, where at large distance two atoms, each prepared in some particular hyperfine state, start to approach each other due to an attractive potential (see Fig.~\ref{Feshbach_mechamism}). At short distances other hyperfine states are coupled in by the interatomic interactions. The set of collision channels is defined by the asymptotic states of the binary system, i.e.~by the hyperfine and angular momentum quantum numbers of the individual atoms that participate in the collision. Due to the Zeeman energy shift, the asymptotic energy $\varepsilon_\alpha+\varepsilon_\beta$ of the atoms and, as a result, also the interaction potentials vary as a function of the magnetic field. In those channels where the total energy $E$ of the atoms is higher than the dissociation threshold, the atoms can escape to infinity and the channel is energetically open. In the channels where at large separations the atoms are below the dissociation threshold, only bound states can form and the channel is energetically closed. In Fig.~\ref{Feshbach_mechamism} schematically only one open and one closed channel are shown.

In the interparticle separation, three different regimes can be distinguished. At long range, where the electron clouds of the atoms do not significantly overlap, the exchange interaction is negligible and the hyperfine interaction dominates. Here $f_{1}$ and $f_{2}$ are good quantum numbers. At short range the exchange interaction dominates, and $S$ and $I$ are good quantum numbers. In the region where the two interactions have the same order of magnitude, $V_{\textrm{ex}} \sim V^{\textrm{hf}}$, the spin coupling changes over a few $\rm a_0$. The atoms are accelerated through this region by the dispersion interaction, and neither $S$ and $I$ nor $f_1$ and $f_2$ are good quantum numbers.

A rigorous way to calculate the wave function of the colliding particles is the coupled channels method. For identical atoms the channel states are defined as
\begin{equation}
|\{\alpha \beta \}^{\pm} ;lm_l \rangle \equiv \frac{|\alpha\rangle_1 |\beta\rangle_2 \pm |\alpha\rangle_2 |\beta\rangle_1}{\sqrt{2(1+\delta_{\alpha \beta})}} \otimes |lm_l \rangle,
\end{equation}
with $|\alpha\rangle$ and $|\beta\rangle$ the single-atom hyperfine states, where the subscripts indicate atom 1 or 2, and $\delta_{\alpha \beta}$ is a Kronecker delta. For distinguishable atoms a direct product $|\alpha\rangle_1 |\beta\rangle_2$ for the internal part is sufficient. The eigenstates of the angular part of the orbital motion $|l, m_l\rangle$ are in a spatial basis represented by spherical harmonics wavefunctions $Y_{l,m_l}(\hat r)$, with $l$ and $m_l$ the relative orbital angular momentum quantum numbers.  The identical particle channel states are (anti-)symmetrized under the exchange of atoms to account for their bosonic or fermionic behavior, and their orbital angular momentum quantization. Solutions $|\Psi\rangle$ to the time-independent Schr\"odinger equation are expanded in terms of the channel states:
\begin{equation}
 \langle {\bf r} | \Psi \rangle =  \sum_{\{\alpha \beta\};l m_{l}} \frac{u_{l m_l\{\alpha \beta\}} (k_{\alpha \beta},r)}{r} i^l Y_{l m_l} (\hat r) |\{\alpha \beta\} \rangle.
\label{CCwavefunction}
\end{equation}
The sum over $\{\alpha \beta\}$ is valid both for distinguishable and identical atoms, however, in the latter case it should be taken over different combinations of $\alpha$ and $\beta$ only.

When we substitute Eq.(\ref{CCwavefunction}) into the Schr\"odinger equation and project onto each of the channel states, we find a set of coupled differential equations for the radial wave functions $u_{lm_l\{\alpha \beta\}}(k_{\alpha \beta},r)$:
\begin{eqnarray}
\left[ -\frac{\hbar^2}{2m} \frac{{\rm d}^{2}}{{\rm d}r^{2}} + \frac{l(l+1)\hbar^{2}}{2m r^2} + \varepsilon_{\alpha} + \varepsilon_{\beta} - E \right] u_{lm_l\{\alpha \beta\}}(k_{\alpha \beta},r) = \nonumber\\
- \sum_{\{\alpha' \beta'\};l'm_{l'}} V^{lm_l,l'm_{l'}}_{\{\alpha \beta\},\{\alpha' \beta'\}}(r) u_{l'm_{l'}\{\alpha' \beta'\}}(k_{\alpha' \beta'},r),
\label{CCequations}
\end{eqnarray}
where the coupling matrix is defined by
\begin{equation}
V^{lm_l,l'm_{l'}}_{\{\alpha \beta\},\{\alpha' \beta'\}}(r) = \langle \{\alpha \beta\};lm_l|\left[ V^{\rm cen} + V^{\rm dd}_{\rm spin}  \right] |\{\alpha' \beta'\};l'm_{l'} \rangle.
\end{equation}
The channel momentum corresponds to the kinetic energy in that channel in the usual way
\begin{equation}
\hbar k_{\alpha \beta} = \sqrt{2m(E-\varepsilon_{\alpha}-\varepsilon_{\beta})},
\label{wavenumber}
\end{equation}
and is positive real for open channels and positive imaginary for closed channels. Together with appropriate boundary conditions that specify the incoming and outgoing waves, the coupled channels equations Eq.(\ref{CCequations}) give a complete description of the scattering process.

To obtain accurate results using the coupled channels method, in principle the interaction potential needs to be known over the full radial range. However, the short-range part of the potential for alkali atoms is often insufficiently known to get the required accuracy, compared to experimental results. The accumulated phase method allows one to replace the inner part of the potential by a simple boundary condition, more details about this procedure can be found for instance in Ref.~\cite{Moerdijk95}.

Solutions to the coupled-channels scattering equations are usually found numerically by solving a system of coupled second-order linear differential equations. However, a formal solution to the time-independent Schr\"odinger equation is given by the Lippmann-Schwinger equation \cite{Sakurai93,Taylor72}
\begin{eqnarray} 
|\Psi^{\pm} \rangle &=& |\chi \rangle + \frac{1}{E^{\pm} - H_0}V|\Psi^{\pm} \rangle \label{LippmannSchwinger1} \\
 &=& |\chi \rangle + \frac{1}{E^{\pm} - H}V|\chi \rangle, 
\label{LippmannSchwinger2} 
\end{eqnarray} 
where $E^\pm =E\pm i\delta$ with $\delta$ approaching zero from positive values. The superscript + (-) indicates outgoing (incoming) spherical wave boundary conditions. The unscattered state  $|\chi \rangle$ is a free solution to $H_0$, which is essentially a sum over the components of the plane wave expansion solution of Eq.~(\ref{CCequations}) with the two-body potential put to zero\footnote{To derive the Lippmann-Schwinger equation (\ref{LippmannSchwinger1}), start from the Schr\"odinger equation $(E-H_0)|\Psi\rangle = V |\Psi\rangle$ and multiply from the left by $(E^{\pm}-H_0)^{-1}$.  It is important that $|\Psi\rangle$ reduces to $|\chi\rangle$ when $V$ vanishes, and therefore the state $|\chi\rangle$, which fulfils $(E-H_0)|\chi\rangle$, is added to the solution. This is allowed, since multiplying the result with $(E-H_0)$ returns the Schr\"odinger equation. To obtain the second form Eq.~(\ref{LippmannSchwinger2}) of the Lippmann-Schwinger equation, start with $(E-H)|\Psi\rangle = 0 = (E-H_0)|\chi\rangle
= (E-H)|\chi\rangle+ V |\chi\rangle$.}. It should be noted that these solutions form a continuum, which are defined only above the collision threshold with kinetic energy $\hbar^2 k^2/2 m = E - \varepsilon_{\rm thr}$. In practice this threshold is the energy of the hyperfine states in which a system of atoms is prepared, and for a collision process this is referred to as the incoming channel. In the remainder of this chapter, we will assume that we only have one open channel, with threshold energy $ \varepsilon_{\rm thr}=\varepsilon_\alpha+ \varepsilon_\beta$ as in Fig.~\ref{Feshbach_mechamism}, which as a result is also the outgoing channel. This means the binary interactions are always elastic, and derived quantities such as scattering amplitudes will be evaluated on the energy shell (``on-shell''). We also limit ourselves to $s$-wave interactions, and therefore we only need to consider the $l=0$ spherical wave component of the plane wave expansion for the free solutions $|\chi \rangle$.

The result of the collision by means of the Lippmann Schwinger solution is commonly expressed in $T$ and $S$-matrices. The on-shell $T$-matrix
\begin{equation}
 T(k)= \langle \chi | \mathcal T | \chi \rangle = \langle \chi |V | \Psi^+ \rangle,
\end{equation}
is related to the transition amplitude of the scattering process, and the on-shell S-matrix
\begin{equation}
 S(k)= \langle \chi | \mathcal S | \chi \rangle = \langle \Psi^{-}| \Psi^{+} \rangle.
 \label{Smatdef}
\end{equation}
These quantities are linked to each other by the relation 
\begin{equation}
 S=1-2\pi i T. 
 \label{STrelation}
\end{equation}
Note that in general the $\mathcal T$ and $\mathcal S$ operators can be evaluated also off-shell. The open and closed channel nature is implicitly hidden in the above expressions, as a result of the implicit spin degrees of freedom. In the next section, we make this more explicit, and this will allow us to link the Feshbach resonance to a bound state in the closed channel.

\section{Feshbach resonances}

Feshbach resonances in two-body collisions are related to the coupling of different spin channels, and can be conveniently described withing the Feshbach projection formalism \cite{Feshbach58,Feshbach62,Feshbach92}. In this approach the total Hilbert space $\mathcal{H}$ describing the spatial and spin degrees of freedom is divided into two subspaces $\mathcal{P}$ and $\mathcal{Q}$. In general $\mathcal{P}$ contains the open channels and $\mathcal{Q}$ the closed channels. The $S$ and $T$ matrices are separated in two parts accordingly. The $P$-part describes the direct interactions in the open-channel subspace, and the $Q$-part describes the effect of the coupling to the closed-channels. Usually the $Q$-part contains the resonances, and the $P$-part is assumed to be non-resonant.

One can construct projection operators $P$ and $Q$, which project onto the subspaces $\mathcal{P}$ and $\mathcal{Q}$, respectively. The Schr\"{o}dinger equation for the two-body collision can then be written as a set of coupled equations: 
\begin{eqnarray} ( E - H_{PP} ) |\Psi_P \rangle & = & H_{PQ} |\Psi_Q \rangle , \label{hpp} \\ ( E - H_{QQ} ) |\Psi_Q \rangle & = & H_{QP} |\Psi_P \rangle .\label{hqq} 
\end{eqnarray} 
Here we use the notation $|\Psi_P \rangle \equiv P|\Psi \rangle$, $|\Psi_Q \rangle \equiv Q|\Psi \rangle$, $H_{PP} \equiv PHP$, $H_{PQ} \equiv PHQ$, $H_{QP} \equiv QHP$, and $H_{QQ} \equiv QHQ$. As already mentioned, we are only interested in scattering processes with only one open channel. The $P$-channel is then simultaneously the incoming and outgoing channel.

We multiply Eq.~(\ref{hqq}) from the left with the resolvent  \cite{Feshbach92} (or Green's) operator $G_{QQ}(E^+) \equiv [E^+ - H_{QQ}]^{-1}$: 
\begin{equation} |\Psi_Q \rangle = \frac{1}{E^+ - H_{QQ}}H_{QP} |\Psi_P \rangle, 
\end{equation} 
where $E^+=E+i\delta$ with $\delta$ approaching zero from positive values. Substituting the expression for $|\Psi_Q \rangle$ in Eq.~(\ref{hpp}), the problem in the $\mathcal{P}$ subspace is equivalent to solving the Schr\"{o}dinger equation 
\begin{equation}
( E - H_{\mathrm{eff}} ) |\Psi_P \rangle = 0,
\label{effschrodinger}
\end{equation}
where the effective Hamiltonian is given by 
\begin{equation} \label{effham} H_{\mathrm{eff}} = H_{PP} + H_{PQ} \frac{1}{E^+ - H_{QQ}} H_{QP}. 
\end{equation} 
The first term in $H_{\mathrm{eff}}$ describes the direct effect of the open-channel subspace $\mathcal{P}$ on the scattering process. The second term in the effective Hamiltonian describes the coupling of $\mathcal{P}$ space to $\mathcal{Q}$ space, propagation through $\mathcal{Q}$ space, and coupling back to $\mathcal{P}$ space again.

The operator $G_{QQ}$ can be expanded in terms of the eigenstates of $H_{QQ}$ (for references see for instance chapter 3.2 of Ref.~\cite{Feshbach92} and more generally the expansion of Green's operator in eigenstates in chapter 8 of Ref.\cite{Taylor72}). These eigenstates will generally have bound states $|\phi_{Q,n} \rangle$ whose eigenvalues $\epsilon_{Q,n}$ form a discrete spectrum, and continuum states $|\phi_Q(\epsilon)\rangle$ whose eigenvalues form a continuous spectrum. These states form an orthonormal set and have relations $\langle\phi_{Q,n} |\phi_{Q,n'} \rangle = \delta_{nn'}$, $\langle\phi_{Q,n} |\phi_Q(\epsilon) \rangle = 0$, and $\langle\phi_Q(\epsilon') |\phi_Q(\epsilon) \rangle = \delta(\epsilon'- \epsilon)$, where $\delta(\epsilon)$ is the Dirac delta function. Then the expansion is given by
\begin{equation}  \frac{1}{E^+ - H_{QQ}} = \sum_n \frac{|\phi_{Q,n} \rangle\langle\phi_{Q,n} |}{E - \epsilon_{Q,n} } + \int \frac{|\phi_Q(\epsilon) \rangle\langle \phi_Q(\epsilon)|}{E^+ - \epsilon} \mathrm{d}\epsilon.
\label{resolventQ}
\end{equation} 
In practice, for ultracold collisions the energy $E$ will be located just above the $\mathcal P$-threshold and sufficiently far away from the $\mathcal Q$-threshold, such that the contribution of the $\mathcal Q$ continuum states can be safely neglected. Moreover, the spacing between the bound states is such that typically only one bound state $|\phi_Q\rangle$ contributes (see also Fig.~\ref{Feshbach_mechamism}), and its eigenvalue $\epsilon_Q$ is close to energy $E$. Then with the substitution of Eq.~(\ref{effham}) and Eq.~(\ref{resolventQ}) into Eq.~(\ref{effschrodinger}), the problem in the $\mathcal{P}$ subspace reduces to 
\begin{equation} ( E - H_{PP} ) |\Psi_P \rangle= H_{PQ} \frac{|\phi_Q \rangle\langle\phi_Q|}{E - \epsilon_Q}H_{QP}|\Psi_P \rangle. 
\end{equation}

Now we can formally solve the coupled problem by multiplying from the left with $G_{PP}(E^+) \equiv [E^+ - H_{PP}]^{-1}$:
\begin{equation} \label{formalP} 
|\Psi_P^+ \rangle= |\phi^+_P \rangle+ \frac{1}{E^+ - H_{PP}}H_{PQ} \frac{ |\phi_Q \rangle\langle \phi_Q |}{E - \epsilon_Q } H_{QP} |\Psi_P^+ \rangle. 
\end{equation} 
The scattering state $|\phi^+_P \rangle$ is a solution of the homogeneous part of Eq.~(\ref{hpp}), $(E-H_{PP})|\phi^+_P \rangle =0$, with outgoing spherical wave boundary conditions \cite{Taylor72}. These scattering states are the equivalent of the continuum states in $\mathcal{Q}$ space and normalized in the same way $\langle \phi_P^+ (E)| \phi_P^+ (E')\rangle = \delta(E-E')$ (defined only for $E>\varepsilon_{\rm thr}$), and are connected to the unscattered states $|\chi(E)\rangle$ via the Lippmann-Schwinger equation (\ref{LippmannSchwinger2}) in $\mathcal{P}$ space:
\begin{equation} |\phi^{\pm}_P \rangle = |\chi \rangle + \frac{1}{E^{\pm} - H_{PP}}V_{PP}|\chi \rangle. \label{LippmannSchwingerP} 
\end{equation} 
Here $V_{PP}=PVP$ is the two-body interaction projected onto the $\mathcal{P}$-channel subspace, and the unscattered states $|\chi \rangle$ in $\mathcal{P}$ space are identical to the free eigenstates of $H_0$.

By making use of the transition matrix due to scattering in the $\mathcal{P}$ subspace only
\begin{equation} 
T_P= \langle \chi|\mathcal{T}_P|\chi \rangle = \langle \chi|V_{PP} |\phi^+_P \rangle = \langle \chi|V |\phi^+_P \rangle, 
\end{equation} 
we find the T-matrix for the total transition amplitude	by multiplying Eq.~(\ref{formalP}) from the left with $\langle \chi | V$, together with relation Eq.~(\ref{LippmannSchwingerP}) to be
\begin{equation} 
T= \langle \chi| V |\Psi_P^+ \rangle =  T_P  + \frac{\langle \phi^-_P|H_{PQ}| \phi_Q \rangle \langle \phi_Q| H_{QP}|\Psi_P^+ \rangle}{E-\epsilon_Q}, 
\end{equation} 
The transition amplitude is now clearly separated in a direct term $T_P$, and a term which results from the coupling to the closed-channel subspace $\mathcal{Q}$. When it comes down to describing Feshbach resonances in terms of measurable quantities, it is more convenient to parametrize the $T$-matrix in quantities related to the uncoupled open and closed channel solutions, {\it i.e.}~coupling matrix elements between open and closed channel solutions. Therefore we have to solve for the component  $\langle \phi_Q| H_{QP}|\Psi_P^+ \rangle$, which we achieve by multiplying Eq.~(\ref{formalP}) from the left by $\langle \phi_Q | H_{QP}$, which results in
\begin{equation}
 \langle \phi_Q| H_{QP}|\Psi_P^+ \rangle = \frac{(E-\epsilon_Q)  \langle \phi_Q| H_{QP}|\phi_P^+ \rangle}{E-\epsilon_Q - \langle \phi_Q |H_{QP} \frac{1}{E^+ - H_{PP}} H_{PQ}|\phi_Q \rangle}.
\end{equation}
With this, the $T$-matrix is expressed by
\begin{equation}
 T= T_P  + \frac{\langle \phi^-_P|H_{PQ}| \phi_Q \rangle \langle \phi_Q| H_{QP}|\phi_P^+ \rangle}{E-\epsilon_Q-A(E)},
\end{equation}
where the term 
\begin{equation} A(E) \equiv \langle \phi_Q |H_{QP} \frac{1}{E^+ - H_{PP}} H_{PQ}|\phi_Q \rangle \label{AE} 
\end{equation} 
in the denominator is the complex energy shift, which will appear to be the energy difference between the bare bound state $|\phi_Q \rangle$ and the dressed \mbox{(quasi-)}bound state of the total coupled system.

From the total $T$-matrix, we can easily go to the total $S$-matrix by using 
relationship Eq.~(\ref{STrelation}). The direct part of the elastic $S$-matrix 
$S_P=\langle \phi^-_P | \phi^+_P \rangle$, that describes the scattering 
process in $\mathcal{P}$ space only, is related in the same way to the direct 
elastic $T$-matrix: $S_P=1-2\pi i T_P$. Since the incoming and outgoing 
spherical wave scattering solutions in $\mathcal{P}$-space are simply related 
via $\langle \phi_P^-| = S_P \langle \phi_P^+ |$, we can write
\begin{equation}
    S =  S_P \left(1 - 2\pi i \frac{\left| \langle \phi_Q | H_{QP} |\phi^+_P \rangle \right|^2}{E - \epsilon_Q - A(E)} \right). 
\label{smatrix1}
\end{equation}

The complex energy shift $A(E)$ can be evaluated by expanding $G_{PP}$ in a complete set of eigenstates of $H_{PP}$, similar to the expansion in closed-channel states of $G_{QQ}$ in Eq.~(\ref{resolventQ}) :
\begin{equation}  \frac{1}{E^+ - H_{PP}} = \sum_n \frac{|\phi_{P,n} \rangle\langle\phi_{P,n} |}{E - \epsilon_{P,n} } + \int_{\varepsilon_{\rm thr}}^{\infty} \frac{|\phi_P^+(\epsilon) \rangle\langle \phi_P^+(\epsilon)|}{E^+ - \epsilon} \mathrm{d}\epsilon.
\label{resolventP}
\end{equation} 
For the closed-channel expansion of $G_{QQ}$, the scattering states could be safely neglected, but this is not the case for $G_{PP}$, where the dominant contribution comes from these scattering states. Therefore $A(E)$ is indeed a complex-valued energy shift, and can be written as $A(E) = \Delta_{\mathrm{res}}(E)-\frac{i}{2}\Gamma(E)$. The real part 
\begin{equation}
 \Delta_{\mathrm{res}}(E) = \sum_n \frac{|\langle \phi_Q| H_{QP} |\phi_{P,n} \rangle |^2}{E - \epsilon_{P,n} } + \wp  \int_{\varepsilon_{\rm thr}}^{\infty} \frac{|\langle \phi_Q| H_{QP}|\phi_P^+(\epsilon) \rangle |^2}{E - \epsilon} \mathrm{d}\epsilon,
 \label{resonanceshiftdefault}
\end{equation}
where $\wp$ refers to the principal value integral, shifts the closed-channel binding energy $\epsilon_Q$, and the imaginary part 
\begin{equation}
 \Gamma(E) = 2\pi |\langle \phi_Q| H_{QP}|\phi_P^+(E) \rangle |^2
 \label{resonancewidthdefault}
\end{equation}
adds a width to the resonance for $E-\varepsilon_{\rm thr}>0$. Note that $A(E)$ is purely real for energies below the $P$-threshold (i.e., $E-\varepsilon_{\rm thr}<0$), and in this regime actually all channels are closed, which means that only bound state solutions can exist. The energy of these dressed bound states can be found by solving for the poles of the $S$ matrix, which will be shown in the next section.

With these definitions, the $S$-matrix can be written as
\begin{equation}
    S =  S_P \left(1 -  \frac{i \Gamma(E) }{E - \epsilon_Q - \Delta_{\mathrm{res}}(E) + \frac{i}{2} \Gamma(E))} \right). 
\label{smatrix}
\end{equation}
This expression, which includes systematically the coupling between open and closed channels, can be generally used to describe single-channel and multi-channel resonance phenomena. For instance, if the resonance position $E_R= \epsilon_Q + \Delta_{\mathrm{res}}(E)$  is assumed to be constant, and if background scattering in the open channel is negligible, the Breit-Wigner cross section Eq.~(\ref{crossection}) is obtained directly from the closed-channel contribution\footnote{The $s$-wave cross section for distinguishable particles \cite{Taylor72} is given by $\sigma_0=4 \pi/k^2 |S-1|^2$.}. On the other hand, a potential resonance, in absence of closed channels, is described by $S_P$ only. For Feshbach resonance physics, all of Eq.~(\ref{smatrix}) needs to be taken into account, in order to derive the magnetic field dependent scattering length Eq.~(\ref{aofB}) and the asymmetric Fano profile of the cross section.

If the background scattering in the open channel is non-resonant, {\it i.e.} there are no nearby bound states in the open channel potential, then $\Delta_{\mathrm{res}}(E)$ in Eq.(\ref{smatrix}) can be assumed constant, and $\Gamma(E)$ to be linearly dependent on $k$. This case of non-resonant background interactions is the most common way to describe Feshbach resonances, and will be worked out in the next section. However, if the $P$-channel has a low-energy potential resonance with a (nearly-)bound state (with $\epsilon_P = -\hbar^2 \kappa^2/2m$) close to threshold, this approximation breaks down.

To circumvent the use of scattering states in $(E^+-H_{PP})^{-1}$ of Eq.(\ref{resolventP}) in favor of expansion into a complete set of discrete states, we expand this propagator to Gamow resonance states, which leads to a Mittag-Leffler expansion \cite{Marcelis04}
\begin{equation}  \label{eq:Mittag}
\frac{1}{E^+-H_{PP}}=\frac{m}{\hbar^2}\sum_{n=1}^\infty\frac{|\Omega_n\rangle\langle\Omega_n^D|}{k_n(k-k_n)},
\end{equation}
where $n$ runs over all poles of the $S_P$ matrix. The Gamow state $|\Omega_n\rangle$ is an eigenstate of $H_{PP}$ with eigenvalue $\epsilon_{P,n}=\hbar^2k_n^2/2m$. Correspondingly, the dual state $|\Omega_n^D\rangle \equiv |\Omega_{n}\rangle^*$, is an eigenstate of $H_{PP}^\dag$ with eigenvalue $\epsilon_{P,n}^*$. These dual states form a biorthogonal set such that $\langle \Omega_n^D\vert \Omega_{n^{\prime }}\rangle=\delta_{nn^{\prime }}$. For bound-state poles $k_n=i\kappa_n$, where $\kappa_n>0$, Gamow states correspond to properly normalized bound states, while for $\kappa_n<0$ these states are unbound, and are also referred to as virtual or resonant states\footnote{Moving a pole in $k$-space from the positive imaginary axis through zero to the negative imaginary axis corresponds to a bound state, moving initially on the so-called first Riemann energy sheet towards the collision threshold at $k=0$. From then on, the bound state disappears from the first Riemann sheet, and moves on as a virtual Gamow state over the second Riemann (or unphysical) sheet \cite{Taylor72}.}.

We assume the scattering in the open channel is dominated by a single bound or virtual state, which means that the $S$-matrix can be described by one simple pole $k_{n}=i\kappa$ in momentum space \cite{Taylor72}. Since the $S$-matrix is unitary, we can write the direct scattering matrix in Eq.~(\ref{smatrix}) as a product of a non-resonant scattering contribution described by an exponential, equivalent to hard-sphere scattering, and a unitary factor that accounts for a pole at $k=i\kappa$: 
\begin{equation}
S_{P}(k)=e^{-2ik r_0}\frac{\kappa-ik}{\kappa+ik}.
\label{SmatrixSP}
\end{equation}
Here $r_0$ is the non-resonant contribution to the $P$-channel scattering length, which is on the order of the range of the interaction potential, given by the van der Waals range $r_{vdW}=(m C_6/8 \hbar^2)^{1/4}$. This non-resonant scattering length $r_0$ is not necessarily equal to the van der Waals range, since both the short-range part of the potential and all the other Gamow states have a residual effect on the non-resonant scattering behavior. The resonant contribution to the $P$-channel scattering length is directly related to the position of the pole, $a_P=1/\kappa$, and the corresponding binding energy $\epsilon_P=-\hbar^2 /(2m a_P^2)$. Note that Eq.~(\ref{SmatrixSP}) by itself is already valid for a single-channel potential resonance description (see also next section).

Since we only consider the bound state pole $\kappa$ with energy $\epsilon
_{P}$ in $\mathcal{P}$ space, as indicated in Fig.~\ref{Feshbach_mechamism}, the Mittag-Leffler series Eq.~(\ref{eq:Mittag}) is reduced to only one term. Therefore, the complex energy shift Eq.~(\ref{AE}) reduces to
\begin{equation}
A(E)=\frac{m }{\hbar ^{2}}\frac{\mathrm{A}}{\kappa
(\kappa+ik)}=\frac{\mathrm{A}}{2|\epsilon_{P}|(1+ika_P)},  \label{eq:ADR}
\end{equation}
where the quadratic coupling matrix element between open-channel Gamow state and closed-channel bound state $\mathrm{A}\equiv \langle \phi _{Q}|H_{QP}|\Omega_{P}\rangle \langle \Omega _{P}^{D}|H_{PQ}|\phi _{Q}\rangle $ is a positive constant.

From this general treatment of the direct scattering process, which now allows also for resonant open channel collisions, we derive from the complex energy shift that the resonance shift is written as
\begin{equation}
\Delta_{\mathrm{res}}(E)=\frac{m}{\hbar^2} \frac{\mathrm{A}}{\kappa^2 + k^2},
\end{equation}
and the corresponding width of the resonance as
\begin{equation}
\Gamma(E)=\frac{m}{\hbar^2} \frac{2 \mathrm{A} k}{\kappa (\kappa^2 + k^2)}.
\end{equation}

\section{Ultracold limit}

For large interatomic separations, the two-body potential will be negligible, and therefore the asymptotic radial wave functions in Eq.~(\ref{CCequations}) of the open channel take the form
\begin{equation}
u_{l}(k,r) \begin{subarray}{c}\phantom{\sim} \\ \displaystyle \sim \\ \scriptstyle{r \rightarrow \infty} \end{subarray} \sin[kr - l\pi /2 + \delta_{l}(k)].
\end{equation}
Compared to the situation where two-body interactions are absent, these solutions are shifted by the scattering phase shift $\delta_{l}(k)$ only. At low collision energies, $s$-wave scattering usually dominates and only the $l=0$ part of the wave function survives. For identical fermions in the same spin-state, the $l=1$ component dominates at low energy, since the wave function has to be anti-symmetric and $s$-wave scattering is forbidden. 

In the remainder of this section, we consider $s$-wave collisions only, and therefore we extract the momentum-dependent $s$-wave scattering phase shift $\delta_0(k)$ from the $S$-matrix expressions that we derived in the previous section. From Eq.~(\ref{Smatdef}) it follows \cite{Taylor72} that $S(k)=\exp (2i\delta_0(k))$.
The scattering length $a_S$ and effective range parameter $R_e$ follow from an expansion up to second order in the wavenumber $k$ of $k \cot \delta_0(k)$.
This results in the well-known effective range expansion \cite{Taylor72}
\begin{equation}
k \cot{\delta_0(k)} = -\frac{1}{a_S} + \frac{R_e}{2}k^2 + \mathcal{O}(k^4).
\label{effrange}
\end{equation}
For a band of energies around threshold, the value of $a_S$ is sufficient to describe the relevant collision physics, however, when $R_e$ is not small, this band of energies is rather narrow, and often one even has to go beyond the effective range expansion. This is particularly true in the presence of (narrow) resonances, and in the following we present more accurate descriptions of the energy-dependent phase shift that are based on the inclusion of the relevant resonance poles. 

\subsection{ Potential resonance}
The s-wave phase shift involving a potential resonance is typical for an open, single channel potential which has a bound or virtual state close to threshold. This behavior is captured by the $S$-matrix in Eq.(\ref{SmatrixSP}), and from the argument of this complex-valued function we find
\begin{equation}
 \delta_0(k)=-k r_0 - \arctan \frac{k}{\kappa} =-k r_0 - \arctan k a_P.
\end{equation}
The corresponding scattering length is then given by
\begin{equation}
 a_S=r_0+a_P.
 \label{pot_res_a}
\end{equation}
After expanding $k \cot{\delta_0(k)}$ to second order in $k$ as in Eq.~(\ref{effrange}), and by eliminating $a_P$ via Eq.~(\ref{pot_res_a}), the effective range coefficient can be written in factors of $1/a_S$ as
\begin{equation}
 R_e = 2 r_0 \left(1- \frac{r_0}{a_S}+\frac{r_0^2}{3 a_S^2} \right),
\end{equation}
which is similar to the expansion that has been derived for a pure long-range van der Waals potential \cite{Flambaum99}.

\subsection{Feshbach resonance with non-resonant open channel interactions} 
When bound or virtual states in the open channel potential are far away from the collision threshold, {\it i.e.}~the open channel scattering is non-resonant, then the resonance shift in Eq.~(\ref{resonanceshiftdefault}) is approximately constant, $\Delta_{\mathrm{res}}(E)\simeq \Delta_{\mathrm{res}}(\varepsilon_{\rm thr}) \equiv \Delta_{\mathrm{res}}$. The resonance energy  $\varepsilon_{\rm res}$ is the difference between the total energy $E$, and the closed-channel bound state $\epsilon_Q$ shifted by $\Delta_{\mathrm{res}}$:
\begin{equation}
\varepsilon_{\rm res} =  E - \epsilon_Q - \Delta_{\mathrm{res}} = \frac{\hbar^2 k^2}{2 m} +\varepsilon_{\rm thr} - \epsilon_Q - \Delta_{\mathrm{res}}.
\label{eres1}
\end{equation}
Since the difference between threshold energy $\varepsilon_{\rm thr}$ and  closed-channel bound state energy $\epsilon_Q$ is approximately linear in  magnetic field around the resonance \cite{Moerdijk95}, we characterize this by $\delta \mu$, the difference in magnetic moment between threshold and bound state. For zero collision energy, the system is exactly on resonance at field value $B_0$, and this allows us to write the resonance energy as
\begin{equation}
\varepsilon_{\rm res} =  \frac{\hbar^2 k^2}{2 m} -\delta \mu (B-B_0).
\label{eres2}
\end{equation}

Without open channel resonances, the energy width $\Gamma(E)=2\pi |\langle \phi_Q | H_{QP} |\phi^+_P(E) \rangle|^2$ is proportional to $k$ in the limit for $k\downarrow 0$, which is a consequence of the $s$-wave scattering state $|\phi_P^+(E)\rangle$ that is propertional to $\sqrt k$, this scaling is also known as Wigner's threshold law \cite{Taylor72}. Therefore we are allowed to write $\Gamma(E) \simeq 2Ck$, with $C$ a constant that characterizes the coupling strength between $P$ and $Q$ \cite{Moerdijk95}. The absence of open-channel resonances also allows us to write the direct part of the $S$ matrix as $S_P (k) = \exp[-2ika_{\mathrm{bg}}]$, where the effect of open-channel interactions is captured by a background scattering length $a_{\mathrm{bg}}$. Then, after substitution of the above expressions in Eq.~(\ref{smatrix}), the resulting expression for the $S$ matrix is given by
\begin{eqnarray} S(k) &=& e^{-2ika_{\mathrm{bg}}} \left( 1-\frac{2iCk}{\varepsilon_{\rm res}  + iCk} \right) \\
&=& e^{-2ika_{\mathrm{bg}}} \left( 1-\frac{2ik}{\frac{\hbar^2}{2 m C} k^2 -\frac{\delta \mu (B-B_0)}{C}  + ik} \right).
\label{singleresonance}
\end{eqnarray} 
The total scattering length derived from this $S$-matrix is then
\begin{equation}
 a_S=a_{\rm bg} - \frac{C}{\delta \mu (B-B_0)},
 \label{non_res_fesh}
\end{equation}
which is equivalent to Eq.~(\ref{aofB}), as the energy width constant can be directly related to the magnetic field width of the resonance via $C=a_{\rm bg} \delta \mu \Delta B$. Associated to the energy width, often a resonance strength length-scale parameter is defined \cite{petrov04} as $R^*=\hbar^2/(2 m \delta \mu \Delta B a_{\rm bg})$, which essentially determines the strength of the $k^2$ term in Eq.~(\ref{singleresonance}). When we insert Eq.~(\ref{non_res_fesh}) into Eq.~(\ref{singleresonance}), and make use of the definition $R^*$, we can write the scattering phase shift as
\begin{equation}
 \delta_0(k)=-k a_{\rm bg} -\arg \left[ R^* k^2+\frac{1}{a_S-a_{\rm bg}} +ik\right].
\end{equation}
The effective range coefficient, again derived using Eq.~(\ref{effrange}) and written in factors of $1/a_S$, is given by
\begin{equation}
 R_e = -2 R^* \left( 1 - \frac{a_{\rm bg}}{R^*} - 
 \frac{2 a_{\rm bg}- \frac{a_{\rm bg}^2}{R^*} }{a_S} + \frac{a_{\rm bg}^2-\frac{a_{bg}^3}{3 R^*}}{a_S^2} \right).
\end{equation}
For narrow Feshbach resonances with non-resonant background interactions, the condition $R^*>>|a_{\rm bg}|$ is satisfied, and therefore close to resonance the effective range parameter is proportional to the resonance strength:
\begin{equation}
 R_e\approx-2 R^*.
\end{equation}

\subsection{Feshbach resonance including resonant open channel interactions}
In the previous section, we also considered a more general description of Feshbach resonances that does allow for resonant states in the open channel. The result was a direct $S_P$-matrix which essentially describes a potential resonance, but in addition the complex energy shift $A(E)$ changes as well by the resonance in $\mathcal P$ space. This more rigorous treatment makes that the background scattering length $a_{\rm bg}$ in the previous paragraph is replaced by two new length scales, which are the potential range $r_0$ and the resonant contribution to the open channel scattering length $a_P$. By writing Eq.~(\ref{eq:ADR}) as
\begin{eqnarray}
A(E)&=& \frac{A}{2|\epsilon_P|} \left(1-\frac{ika_P}{1+ika_P}\right) \\
&=&\frac{A}{2|\epsilon_P|}-\frac{\hbar^2}{2 m R^*} \frac{ik}{1+ika_P},  \label{eq:ADR2}
\end{eqnarray}
we relate the coupling strength again to a resonance strength parameter $R^*=\hbar^2 |\epsilon_P|/(m A a_P)=\hbar^2/(2 m \delta \mu \Delta B a_P)$, which is now inversely proportional to $a_P$ (instead of $a_{\rm bg}$). Now we substitute Eq.~(\ref{eq:ADR2}) and Eq.~(\ref{SmatrixSP}) in the $S$-matrix Eq.~(\ref{smatrix}), and by making use of Eq.~(\ref{eres2}) with the identification $\Delta_{\rm res}(\varepsilon_{\rm thr})=A/2|\epsilon_P|$, we derive the total scattering length
\begin{equation}
 a_S=r_0+a_P \left(1-\frac{\Delta B}{B-B_0} \right),
\end{equation}
and the scattering phase shift which is given by
\begin{equation}
 \delta_0(k)=-k r_0 - \arctan k a_P -\arg \left[ R^* k^2+\frac{1}{a_S-r_0-a_P} +\frac{ik}{1+ik a_P}\right].
\end{equation}
From this phase shift we are able to extract the effective range coefficient, written in factors of $1/a_S$:
\begin{equation}
 R_e = -2 R^* \left( 1 - \frac{r_0}{R^*} - 
 \frac{2(r_0+a_P)- \frac{r_0^2}{R^*} }{a_S} + \frac{(r_0+a_P)^2-\frac{r_0^3}{3 R^*}}{a_S^2} \right).
\end{equation}
We find that $R_e$ changes rapidly with $a_P$, notably via $R^* \sim a_P^{-3}$. Therefore control over the width of the Feshbach resonance can be achieved if one is able to tune the position of the resonant state in the open channel. This can for instance be achieved by applying strong electric fields~\cite{Marcelis08}.

The poles of the total $S$ matrix Eq.~(\ref{smatrix}) now give rise to dressed bound states and dressed virtual states of the coupled system. For finding these poles, one has to solve the equation
\begin{equation}
(k-i\kappa)\left( E-\epsilon _{Q}-A(E)\right) =0
\label{eq:PoleEqn}
\end{equation}for $k$, which, if worked out including the parameters that characterize the resonance, can be written as
\begin{equation}
 (1+ik a_P) \left(R^* k^2+\frac{1}{a_S-r_0-a_P}\right) +ik =0.
\end{equation}
Just below threshold, in the regime where the total scattering length $a_S$ is much larger then all other length scales in the problem, this equation yields the characteristic solution for the dressed binding energy
\begin{equation}
E= \frac{\hbar^2 k^2}{2 m} \approx-\frac{\hbar^2}{2m a_S^2}.
\end{equation}
This clearly indicates that exactly on resonance, the dressed bound state becomes degenerate with the collision threshold. The identification of Feshbach resonance positions with degenerate bound states makes it possible to  analyse Feshbach spectroscopic data using discrete bound state models \cite{tiecke10}.

\section{Conclusions}

In this chapter, we discussed several scattering resonance phenomena in the context of ultracold gases. We discussed the different nature of single-channel (shape and potential) resonances and multi-channel (Feshbach and Fano) resonances. The different contributions to the interparticle interaction have been discussed, and a treatment of the Feshbach resonance formalism has been given. This formalism has been generalized to account also for resonant background interactions, {\it i.e.} for systems where the background scattering length is much larger than the range of the interaction potentials.
\bibliography{ChapterKokkelmans}

\begin{thebibliography}{25}
\providecommand{\natexlab}[1]{#1}
\providecommand{\url}[1]{\texttt{#1}}
\expandafter\ifx\csname urlstyle\endcsname\relax
  \providecommand{\doi}[1]{doi: #1}\else
  \providecommand{\doi}{doi: \begingroup \urlstyle{rm}\Url}\fi

\bibitem{Taylor72}
J.~R. Taylor, \emph{Scattering Theory}. (Robert E. Krieger Publishing Company,
  1987), 3rd edition.

\bibitem{Sakurai93}
J.~J. Sakurai, \emph{Modern Quantum Mechanics (Revised Edition)}. (Addison
  Wesley, 1993), 1 edition.

\bibitem{Kukulin89}
V.~I. Kukulin, V.~M. Krasnopol'sky, and J.~Hor\'acek, \emph{Theory of
  Resonances}. (Kluwer Academic Publishers, 1989).

\bibitem{Feshbach92}
H.~Feshbach, \emph{Theoretical Nuclear Physics}. (John Wiley and Sons, Inc,
  1992).

\bibitem{Feshbach58}
H.~Feshbach, Unified theory of nuclear reactions, \emph{Ann. of Phys.} {\bf 5},
  \penalty0 357,  (1958).

\bibitem{Feshbach62}
H.~Feshbach, Unified theory of nuclear reactions {II}, \emph{Ann. of Phys.}
  {\bf 19}, \penalty0 287,  (1962).

\bibitem{stwalley76}
W.~C. Stwalley, Stability of spin-aligned hydrogen at low temperatures and high
  magnetic fields: New field-dependent scattering resonances and
  predissociations, \emph{Phys. Rev. Lett.} {\bf 37}, \penalty0 1628--1631,
  (1976).

\bibitem{chin10}
C.~Chin, R.~Grimm, P.~Julienne, and E.~Tiesinga, Feshbach resonances in
  ultracold gases, \emph{Rev. Mod. Phys.} {\bf 82}, \penalty0 1225--1286,
  (2010).

\bibitem{tiesinga93}
E.~Tiesinga, B.~J. Verhaar, and H.~T.~C. Stoof, Threshold and resonance
  phenomena in ultracold ground-state collisions, \emph{Phys. Rev. A}. {\bf
  47}, \penalty0 4114--4122,  (1993).

\bibitem{inouye98}
S.~Inouye, M.~R. Andrews, J.~Stenger, H.-J. Miesner, D.~M. Stamper-Kurn, and
  W.~Ketterle, Observation of {F}eshbach resonances in a bose-einstein
  condensate, \emph{Nature}. {\bf 392}, \penalty0 151,  (1998).

\bibitem{courteille98}
P.~Courteille, R.~S. Freeland, D.~J. Heinzen, F.~A. van Abeelen, and B.~J.
  Verhaar, Observation of a {F}eshbach resonance in cold atom scattering,
  \emph{Phys. Rev. Lett.} {\bf 81}, \penalty0 69--72,  (1998).

\bibitem{Kokkelmans01}
S.~J. J. M.~F. Kokkelmans, H.~M.~J. Vissers, and B.~J. Verhaar, Formation of a
  {B}ose condensate of stable molecules via a {F}eshbach resonance, \emph{Phys.
  Rev. A}. {\bf 63}, \penalty0 031601,  (2001).

\bibitem{Lang08}
F.~Lang, K.~Winkler, C.~Strauss, R.~Grimm, and J.~H. Denschlag, Ultracold
  triplet molecules in the rovibrational ground state, \emph{Phys. Rev. Lett.}
  {\bf 101}, \penalty0 133005 (Sep, 2008).

\bibitem{Ni08}
K.-K. Ni, S.~Ospelkaus, M.~H.~G. de~Miranda, A.~Pe’er, B.~Neyenhuis, J.~J.
  Zirbel, S.~Kotochigova, P.~S. Julienne, D.~S. Jin, and J.~Ye, A high
  phase-space-density gas of polar molecules, \emph{Science}. {\bf 322},
  \penalty0 231,  (2008).

\bibitem{Fano61}
U.~Fano, Effects of configuration interaction on intensities and phase shifts,
  \emph{Phys. Rev.} {\bf 124}, \penalty0 1866--1878,  (1961).

\bibitem{Smirnov65}
B.~Smirnov and M.~Chibisov, Electron exchange and changes in the hyperfine
  state of colliding alkaline metal atoms, \emph{JETP}. {\bf 21}, \penalty0
  624,  (1965).

\bibitem{Moerdijk95}
A.~J. Moerdijk, B.~J. Verhaar, and A.~Axelsson, Resonances in ultracold
  collisions of $^{6}\mathrm{Li}$, $^{7}\mathrm{Li}$, and $^{23}\mathrm{Na}$,
  \emph{Phys. Rev. A}. {\bf 51}, \penalty0 4852--4861,  (1995).

\bibitem{marte02}
A.~Marte, T.~Volz, J.~Schuster, S.~D\"urr, G.~Rempe, E.~G.~M. van Kempen, and
  B.~J. Verhaar, Feshbach resonances in rubidium 87: Precision measurement and
  analysis, \emph{Phys. Rev. Lett.} {\bf 89}, \penalty0 283202,  (2002).

\bibitem{chin04}
C.~Chin, V.~Vuleti\ifmmode~\acute{c}\else \'{c}\fi{}, A.~J. Kerman, S.~Chu,
  E.~Tiesinga, P.~J. Leo, and C.~J. Williams, Precision {F}eshbach spectroscopy
  of ultracold ${\mathrm{cs}}_{2}$, \emph{Phys. Rev. A}. {\bf 70}, \penalty0
  032701,  (2004).

\bibitem{messiah65}
A.~Messiah, \emph{Quantum Mechanics}. (North Holland Publishing Company, 1967).

\bibitem{Marcelis04}
B.~Marcelis, E.~G.~M. van Kempen, B.~J. Verhaar, and S.~J. J. M.~F. Kokkelmans,
  Feshbach resonances with large background scattering length: Interplay with
  open-channel resonances, \emph{Phys. Rev. A}. {\bf 70}, \penalty0 012701,
  (2004).

\bibitem{Flambaum99}
V.~V. Flambaum, G.~F. Gribakin, and C.~Harabati, Analytical calculation of
  cold-atom scattering, \emph{Phys. Rev. A}. {\bf 59}, \penalty0 1998--2005,
  (1999).

\bibitem{petrov04}
D.~S. Petrov, Three-boson problem near a narrow {F}eshbach resonance,
  \emph{Phys. Rev. Lett.} {\bf 93}, \penalty0 143201,  (2004).

\bibitem{Marcelis08}
B.~Marcelis, B.~Verhaar, and S.~Kokkelmans, Total control over ultracold
  interactions via electric and magnetic fields, \emph{Phys. Rev. Lett.} {\bf
  100}, \penalty0 153201,  (2008).

\bibitem{tiecke10}
T.~G. Tiecke, M.~R. Goosen, J.~T.~M. Walraven, and S.~J. J. M.~F. Kokkelmans,
  Asymptotic-bound-state model for {F}eshbach resonances, \emph{Phys. Rev. A}.
  {\bf 82}, \penalty0 042712,  (2010).

\end{thebibliography}
\end{document}